\documentclass[aps,prl,twocolumn,english,superscriptaddress,longbibliography,nobibnotes]{revtex4-2}
\usepackage{graphicx}
\usepackage[usenames,dvipsnames]{color}
\usepackage{bbm}
\usepackage{bm}
\usepackage{amsmath}
\usepackage{amssymb}
\usepackage{mathptmx}
\usepackage[colorlinks=true,linkcolor=Blue,citecolor=Blue,urlcolor=Blue]{hyperref}
\usepackage[version=3]{mhchem}

\graphicspath{{figures/compressed/}}

\begin{document}

\title{Pre-thermalization via self-driving and external driving of extensive subsystems}

\author{Finn Lasse Buessen}
\thanks{These authors contributed equally to this work}
\affiliation{Department of Physics, University of Toronto, Toronto, Ontario M5S 1A7, Canada}

\author{Hyun-Yong Lee}
\thanks{These authors contributed equally to this work}
\affiliation{Department of Applied Physics, Graduate School, Korea University, Sejong 30019, Korea}
\affiliation{Division of Display and Semiconductor Physics, Korea University, Sejong 30019, Korea}
\affiliation{Interdisciplinary Program in E$\cdot$ICT-Culture-Sports Convergence, Korea University, Sejong 30019, Korea}

\author{Tarun Grover}
\affiliation{Department of Physics, University of California at San Diego, La Jolla, California 92093, USA}

\author{Yong Baek Kim}
\email{ybkim@physics.utoronto.ca}
\affiliation{Department of Physics, University of Toronto, Toronto, Ontario M5S 1A7, Canada}

\begin{abstract}
We investigate the non-equilibrium states of an interacting multi-component quantum system when only an extensive subsystem is quantum-quenched or driven from the ground state. 
As a concrete example, we consider a system where two XXZ spin chains are coupled to a transverse field Ising (TFI) chain, and only the transverse field in the TFI chain is quantum-quenched or periodically driven in time, starting from an initially ordered state. 
This system is studied using density matrix renormalization group (DMRG) simulations and various entanglement entropy diagnostics. 
In the case of quantum quenching, when the transverse field is suddenly switched on to become the largest energy scale, the resulting internal dynamics leads to a pre-thermal steady state with persistent oscillating magnetization (`self-driving') and emergent conservation laws. 
Upon applying the time-dependent drive to the TFI chain (`external driving'), sufficiently fast drive gives rise to a pre-thermal steady state with finite magnetization, whereas a slow drive generates a high-temperature disordered state. We briefly discuss the experimental implementation of our protocol in organic materials with quantum-tunneling hydrogen atoms.
\end{abstract}

\date{\today}
\maketitle


{\it Introduction --}
Our ability to understand non-equilibrium quantum states of interacting quantum matter would significantly expand the scope of accessible quantum phases of condensed matter and cold atom systems~\cite{Eisert2015,Oka2019,Rudner2020}. 
Of particular interest are the pre-thermal states that may persist for an exponentially long period of time and arise as a consequence of emergent quasi-conservation laws. 
For example, in closed systems, the Floquet-type (periodic in time) drive may, in general, heat up the system to the infinite temperature state~\cite{Lazarides2014,Kim2014,DAlessio2014}; however, the integrable systems~\cite{Lazarides2014a,Gritsev2017} and many-body localized (MBL) systems~\cite{Lazarides2015,Ponte2015,Rehn2016} show pre-thermal steady states in accordance with associated conserved quantities. 
Alternatively, quantum quenching in certain systems may lead to quasi-conserved quantities, which can then give rise to long-lived pre-thermal steady states.  
Yet, most of the previous studies are limited to cases in which the entire system is quenched or driven together (see e.g.~\cite{cayssol2013floquet, bukov2015universal, eckardt2017colloquium, oka2019floquet, harper2020topology}). 
Many condensed matter and cold atomic systems, however, consist of multiple degrees of freedom, and different parts of the system may possess distinct natural time scales. 
In this setting, one may ask whether quenching or driving only a subsystem would necessarily heat up the entire system or whether there may be a different limit in which non-trivial dynamic states can be realized. 
Moreover, we may also ask how one could effectively characterize the non-equilibrium states in such subsystem-driven interacting quantum systems.        

\begin{figure}[t]
	\includegraphics[width=\linewidth]{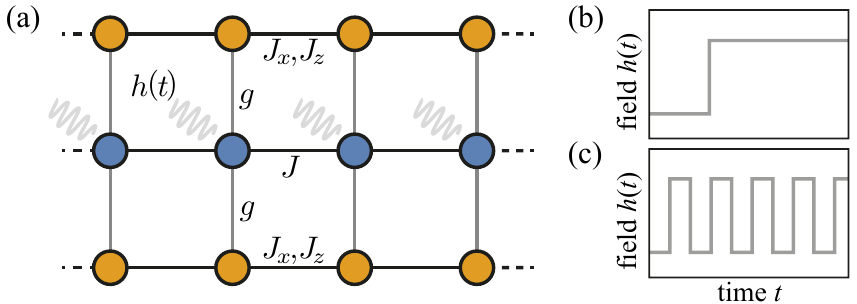}
	\caption{{\bf Coupled three-chain model}. (a)~The top and bottom chains (yellow sites) are XXZ models with exchange constants $J_x$ and $J_z$. The middle chain (blue sites) is a TFI chain with interaction $J$ and on-site time-dependent transverse field $h(t)$. The chains interact via a term $g \sigma_i^z (\sigma_{i,t}^z-\sigma_{i,b}^z)$, see text for details. (b)~Quench protocol with a sudden onset of the transverse field. (c)~Periodic driving protocol.}
	\label{fig:model}
\end{figure}

In this work, we consider an interacting spin model, where two XXZ spin chains are coupled to a transverse field Ising (TFI) spin chain and only the latter is quenched or driven by a time-dependent transverse field, see the schematic illustration in Fig.~\ref{fig:model}. 
This model is partially motivated by the theory of organic materials in which layers of electronic degrees of freedom are coupled via hydrogen bonds; quantum tunneling between the bi-stable ground state configurations of the hydrogen bonds can then be modeled by transverse-field Ising spins~\cite{Yamamoto2016, Naka2018}.
Using the density matrix renormalization group (DMRG), we study both equilibrium and non-equilibrium scenarios in our model. 
First, we establish the equilibrium phase diagram as a function of the time-independent transverse field strength in the TFI chain and the exchange interaction scales in the system. 
It is demonstrated that there exists a direct transition from an ordered state (ordered in both the XXZ and TFI subsystems) to a fully quantum disordered state, where neither of the subsystems shows finite magnetization. 
We then consider a quantum-quench protocol where a large transverse field in the TFI chain is suddenly turned on, starting from an ordered ground state. 
It is found that when the transverse field becomes the largest intrinsic energy scale, the entire system enters a pre-thermal state with oscillating magnetization in both the XXZ and TFI chains. 
We explain this phenomena by constructing an effective Hamiltonian where emergent decoupling between the XXZ and TFI chains occurs and new quasi-conservation laws arise. 
Next, by imposing a periodic drive on the transverse field in the TFI chain and starting from an ordered ground state of the whole system, we investigate how the resulting non-equilibrium state evolves as a function of the driving frequency. 
We demonstrate that the low and high frequency drives of the TFI subsystem lead to very different behaviors of the composite model. 
We show that if the drive frequency is sufficiently large, the system may remain in a long-lived, symmetry-broken pre-thermal regime and maintain its finite magnetization. 
Remarkably, the polarizable environment of XXZ chains significantly enhances the stability of the pre-thermal regime when compared to an isolated TFI chain.


{\it Model --}
We investigate a model of three coupled spin chains that consists of a transverse field Ising (TFI) chain at the center and two XXZ spin chains at the top~($t$) and bottom~($b$) of the system as illustrated in Fig.~\ref{fig:model}. 
The Hamiltonian of the full system reads $H = H_\mathrm{TFI} + H_\mathrm{XXZ,t} + H_\mathrm{XXZ,b} + H_\mathrm{int}$, where the TFI and XXZ terms are given by
\begin{align}
& H_\mathrm{TFI} = J \sum\limits_{\langle i,j\rangle} \sigma^z_i \sigma^z_j + h(t) \sum\limits_i \sigma_i^x \,, \nonumber\\
& H_{\mathrm{XXZ},\alpha} = \sum\limits_{\langle i,j\rangle} J_x \left( \sigma_{i,\alpha}^x \sigma_{j,\alpha}^x + \sigma_{i,\alpha}^y \sigma_{j,\alpha}^y \right) + J_z \sigma_{i,\alpha}^z \sigma_{j,\alpha}^z \,.
\end{align}
Here $\langle i,j \rangle$ denote nearest neighbor lattice sites $i$ and $j$ within each chain, $\alpha=t,b$ discriminates top and bottom chain, and $\sigma^{\gamma}$ are the Pauli matrices with $\gamma=x,y,z$. 
The three spin chains are locally coupled via an interaction term $H_\mathrm{int} = g \sum_i \sigma_i^z ( \sigma_{i,t}^z - \sigma_{i,b}^z )$, which couples the local magnetization $\sigma_i^z$ of the TFI chain to the local magnetization difference of the two XXZ chains via the interaction constant~$g$. 
In addition to the symmetries corresponding to the $U(1)$ rotation about the z-axis separately in the top and bottom chains, there exists also a global Ising symmetry corresponding to $ \sigma_i^z \mapsto  -\sigma_i^z, \sigma_{i,\alpha}^z \mapsto - \sigma_{i,\alpha}^z$.
Note that the top and bottom XXZ chains do not interact directly. 
We fix $J=1$, $J_x=1$, $g=0.5$, and investigate the role of varying $h(t)$ and $J_z$ in this Letter.

\begin{figure}
	\includegraphics[width=\linewidth]{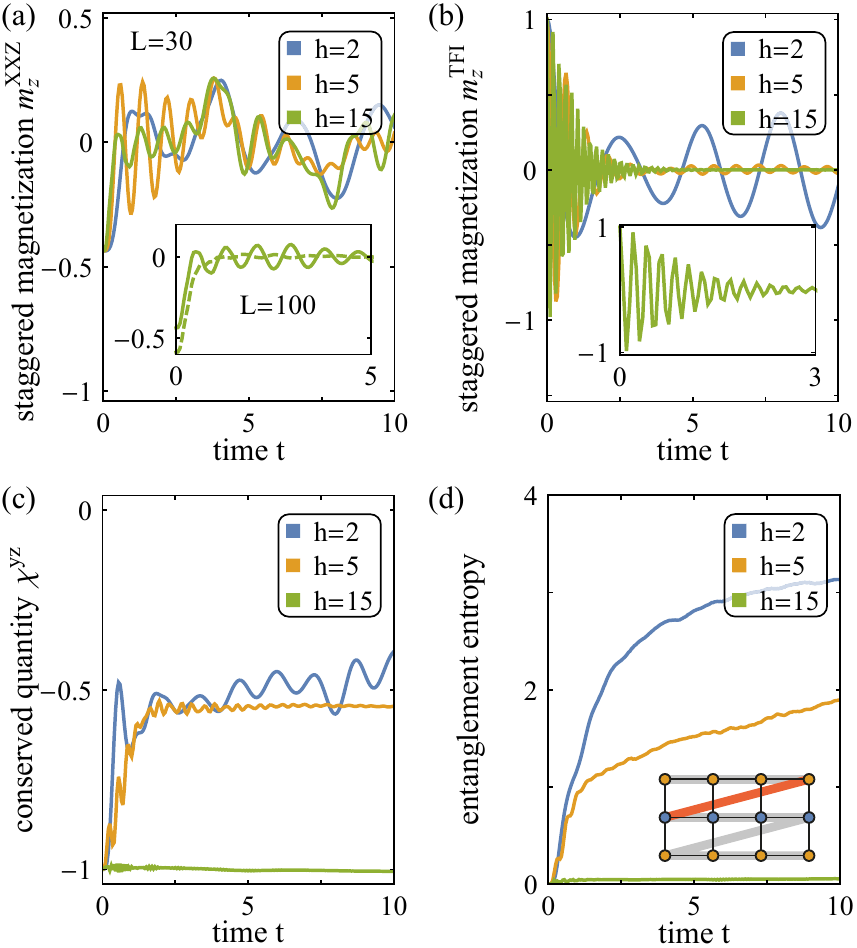} 
	\caption{{\bf Quench dynamics} after a sudden shift in the magnetic field from $h=0$ to various values $h=2,5,15$. Data shown is for finite chain length $L=30$ with periodic boundary conditions, interaction $J_z=0$, and maximum bond dimension $\chi_\mathrm{max}=512$. (a)~Evolution of the staggered magnetization in the XXZ subsystem. The inset shows data for larger system size $L=100$. The dashed line depicts data for $J_z=0.5$. (b)~Staggered magnetization in the TFI chain. The inset shows the short-time dynamics for $h=15$. (c)~Conserved quantity $\chi^{yz}$ in the TFI chain. (d)~Bipartite entanglement entropy at the cut between the (top) XXZ chain and the remainder of the system; this data is for $L=10$ and $\chi_\mathrm{max}=1024$. The inset illustrates the layout of the MPS by thick gray lines, the bipartitioning cut is colored red. }
	\label{fig:DMRG_quench_composite}
\end{figure}

{\it Method --} 
Our calculations of the equilibrium ground state are based on the density matrix renormalization group (DMRG) method~\cite{White1992,White1993,McCulloch2008}. 
For this purpose, unless indicated otherwise, the three-chain model is mapped onto a one-dimensional matrix product state (MPS) by winding the MPS along the first XXZ chain, then the TFI chain, and finally along the second XXZ chain (see inset of Fig.~\ref{fig:DMRG_quench_composite}d). 
To study the quantum quench or periodic driving from the ground states, we employ the time dependent variational principle (TDVP)~\cite{Jutho2011}. 
For characterizing the entanglement between different subsystems, in addition to more conventional measures, we calculate the `quantum disentanglement liquid' (QDL) diagnostic~\cite{Tarun2014, garrison2017partial, veness2017quantum, Tarun2020}. 
The QDL diagnostic is designed to extract an effective entanglement between two subsystems of a tripartite system: We consider the whole system as a union of three subsystems $A$, $B$, and $C$ and project the subsystem $C$ into a given basis state, $X\equiv\{x_c\}$, resulting in a state $|\psi_{AB}^X \rangle \equiv \bigotimes_{c\in C}\langle x_c|\psi\rangle$. 
Then, the QDL diagnostic is defined as $S_{\rm QDL} \equiv \sum_{X} p_X S_{AB}^X$, where $p_X \equiv \langle\psi_{AB}^X|\psi_{AB}^X\rangle$, and $S_{AB}^X\equiv -\mathrm{Tr}(\rho_{A}^X \log \rho_{A}^X)$ is the entanglement entropy~(EE) of the reduced density matrix $\rho_{A}^X \equiv \mathrm{Tr}_B (|\psi_{AB}^X \rangle\langle\psi_{AB}^X|)$. 
Consequently, the QDL diagnostic reflects an effective entanglement between subsystems $A$ and $B$, and as shown in Ref.~\cite{Tarun2020} it also bounds the conditional entanglement between the subsystems. 
See the supplemental material (SM) for further details on the QDL diagnostic.


{\it Ground state phase diagram --}
We first carve out the ground state phase diagram of the three-chain model as a function of $h\equiv h(t)$ and $J_z$. 
We observe two possible ground state phases of the three-chain model. 
At small values for the transverse field $h$~$(<J)$, or large values of $J_z$~$(>J_x)$, the ground state is a composite magnetic order where the global Ising symmetry is broken and all three chains obtain a finite magnetization along the z-axis. 
In the opposite limit, i.e., $h>J$ and $J_z<J_x$, the magnetization vanishes across all three chains as they remain disordered; note that in this parameter regime the chains would also be disordered in the absence of any inter-chain interaction, i.e., $g=0$, where the XXZ (TFI) chains form a Luttinger liquid (paramagnet)~\cite{Pfeuty1970,Fogedby1978,Alcarez1995,Kudasov2018}. 
The detailed phase diagram is discussed in the SM. 
We note that our ground state phase diagram has strong resemblance to that for models of organic materials such as $\kappa$-H$_3$(Cat-EDT-TTF)$_2$ \cite{Yamamoto2016, Naka2018}, whereby the XXZ chains play the role of electronic spins, while the TFI chain spins play the role of hydrogen atoms tunneling in a double-well potential.


{\it `Self-driven' pre-thermalization via quantum quenching --} 
We imagine a scenario where we start from an ordered ground state of the entire system at vanishing transverse field in the TFI chain and then suddenly change the transverse field to become finite -- a so-called quantum quench (schematically depicted in Fig.~\ref{fig:model}c). 
In this setting we shall investigate the time evolution of the magnetization in the XXZ chains for various strengths of the transverse field.
While for small transverse field strength the time evolution is expected to be chaotic and therefore difficult to predict in detail, we are able to formulate an analytically reasoned expectation for the case when the transverse field is much larger than all other interaction scales in the system. 
For our analysis we employ the following mapping~\cite{abanin2017rigorous, Mori2018}. 
Let us write $H = h \sum_i \sigma_i^x +V$ and utilize the interaction picture with $V$ treated as a perturbation to the transverse-field term. 
The interaction picture many-body wave function $|\psi^{I}(t)\rangle$ is given by $i \frac{d |\psi^{I}(t) \rangle}{dt} = V^{I} (t) |\psi^{I}(t)\rangle$, where $V^{I}(t) = U(t) V U(t)^{\dagger}$ and $U(t) = \exp(i  t h \sum_i \sigma_i^x)$. 
As one may readily check, $U(t + 2\pi/h) = U(t)$ and thus $V^I(t)$ is \textit{time-periodic} with frequency $\omega\equiv T^{-1}=h/2\pi$, despite $H$ not having any such periodicity. 
One can now borrow the results on pre-thermalization in time-periodic systems \cite{abanin2017rigorous, Mori2016,Abanin2017,Mori2018} to understand the behavior of $|\psi^I(t)\rangle$. 
The essential point is that when $h$ is much larger than all other intrinsic energy scales in the problem, then for times that are exponential in $h$, the interaction picture wave function $|\psi^I(t)\rangle$ will evolve with an effective, time-independent Hamiltonian $V_\mathrm{eff} = \overline{V^I(t)}$ that equals the time-averaged $V^I(t)$. 
Explicitly, $|\psi^I(t) \rangle = e^{-i V_{\rm eff}\, t} |\psi^I(0)\rangle$. 
One can then obtain the time-dependence of any observable $O$ using the relation $\langle O\rangle (t) = \langle \psi^I(t)|O^I(t)|\psi^I(t)\rangle$. 
For our problem, we obtain $V_\mathrm{eff} = H_\mathrm{XXZ,t} + H_\mathrm{XXZ,b} + \frac{J}{2} \sum_{\langle i,j\rangle} ( \sigma^y_i \sigma^y_j + \sigma^z_i \sigma^z_j )$.
Remarkably, in this effective description, the  three chains decouple into three separate integrable systems, and further there now exists an emergent $U(1)$ symmetry in the TFI chain which corresponds to arbitrary rotations around the $x$-axis. 

We are now equipped to calculate the local magnetization in the (bottom) XXZ chain and in the TFI chain. 
We anticipate a qualitative distinction between the XXZ and TFI magnetizations: 
The local magnetization operator in the XXZ chain $\sigma^{z,I}_{b}(t) = U \sigma^{z}_{b} U^{\dagger} = \sigma^{z}_{b} $ is independent of time since $U$ and $\sigma^{z}_{b}$ commute, while for the TFI chain it is explicitly time dependent, $\sigma^{z,I}(t) = \sigma^{z} \cos(2 h t) + \sigma^{y} \sin(2 ht)$ (note that we suppressed the site labels to improve readability). 
Therefore, the local magnetization in the XXZ chain $\langle \sigma^{z}_{b} \rangle$ is fully determined by the quench dynamics in an integrable XXZ chain while in the TFI chain $\langle \sigma^{z} \rangle$ will exhibit oscillations with period $T=\pi/h$ in addition to its behavior determined by the TFI chain quench dynamics. 
As an example, we consider the case $J_z = 0$. 
Using results from Ref.~\cite{barmettler2010quantum}, one finds $\langle \sigma^{z}_{b}\rangle (t) \approx \cos(8 J_x t - \pi/4)/\sqrt{t}$ for $ t < c L$, where $c$ is a constant. 
For $t > c L$, the finite-size effects take over and lead to oscillations whose time-period is proportional to the finite-size gap $\sim 8 \pi J_x/L$. 
Indeed, in our numerical simulations we observe systematic oscillations in the staggered magnetization $m_z^\mathrm{XXZ}\equiv\frac{1}{L}\sum_i (-1)^i \sigma^z_{i,t}$ of the XXZ subsystem with period $T\approx \pi/(4J_x)$, see Fig.~\ref{fig:DMRG_quench_composite}a, which corresponds to $L = 30$. 
Since oscillations are cut off by finite-size effects at time $O(L)$, we also studied much larger system size, $L=100$, and found agreement with the prediction that $T\approx \pi/(4J_x)$ -- see the inset of Fig.~\ref{fig:DMRG_quench_composite}a. The same inset also shows the effect of including non-zero $J_z = 0.5$, in which case it is expected that the magnetization decays exponentially, modulated with weak oscillatory behavior~\cite{barmettler2010quantum}, in line with our observation.

Turning next to the TFI chain, we find that the staggered magnetization $m_z^\mathrm{TFI}\equiv\frac{1}{L}\sum_i (-1)^i \sigma^z_{i}$ oscillates with period $T=\pi/h$ in agreement with our prediction (Fig.~\ref{fig:DMRG_quench_composite}b).
Two additional predictions of the pre-thermal physics encapsulated in $V_{\rm eff}$ can be made: (i)~the emergence of the conserved quantities $\chi^{yz}\equiv\frac{1}{L}\sum_{\langle i,j \rangle} (\sigma^y_i \sigma^y_j + \sigma^z_i \sigma^z_j)$ and $m_x^\mathrm{TFI}\equiv \frac{1}{L} \sum_i \sigma_i^x$, and (ii)~the decoupling of the three chains. 
We have verified numerically that the quantity $\chi^{yz}$ remains constant after a quench to strong transverse field $h=15$, as depicted in Fig.~\ref{fig:DMRG_quench_composite}c. 
Similarly, we also observe the conservation of $m_x^\mathrm{TFI}$ (see SM). 
To detect the decoupling of the three chains, we studied the entanglement between the top XXZ chain and the remainder of the system, and find that it remains constant for sufficiently strong transverse field, see Fig.~\ref{fig:DMRG_quench_composite}d. 
By symmetry, the entanglement between the bottom chain and the remainder of the system shows the same behavior, implying a decoupling of all three chains.


{\it Extensive subsystem drive --}
We now consider a periodic step-like transverse field $h(t)$ akin to the driving protocol illustrated in Fig.~\ref{fig:model}c. 
During a single period $T$, we chose $h(t) = h_\mathrm{max}$ for $t < T/2$ and $h(t) = 0$ for $T/2 < t < T$. 
A translationally invariant closed quantum system subject to an external drive is generally expected to heat to infinite temperature in the long-time limit. 
A slow drive is indeed generally associated with fast heating. 
In contrast, the heating rate can become exponentially small in rapidly driven systems, resulting in long-lived pre-thermal states~\cite{abanin2017rigorous, Mori2016,Abanin2017,Mori2018}. 
Such pre-thermalization behavior is typically studied when the whole system or a non-extensive subsystem is driven externally, and the case of an \textit{extensive} subsystem drive has not received much attention. 
Here, we address the question of whether an extensive subsystem drive can also be associated with pre-thermal behavior and study how the system evolves under the slow ($\omega=0.5$) and fast ($\omega=5$) subsystem drives with $h_\mathrm{max} = 1.5$. 
We first consider the case when the system is initially prepared in an eigenstate of the time-averaged Hamiltonian, i.e., it is an eigenstate of the equilibrium model with $h(t) \equiv h_\mathrm{max} /2$. 
We keep track of three quantities during the time evolution: the magnetization, the bond-dependence of the bipartition EE, and the QDL diagnostic for the effective entanglement between a single XXZ chain and the TFI chain. 

\begin{figure}
	\includegraphics[width=\linewidth]{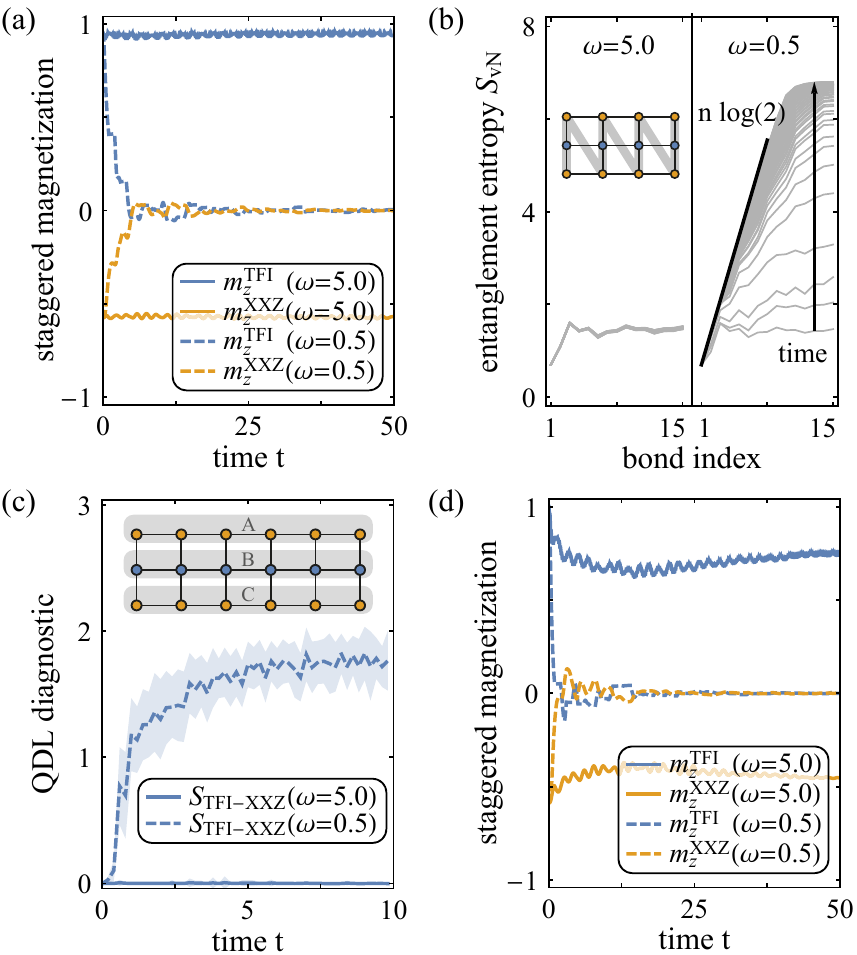} 
	\caption{{\bf Time evolution} of the system under fast ($\omega=5$) and slow ($\omega=0.5$) periodic driving. Data shown is for $L=30$ and $\chi_\mathrm{max}=256$. (a)~Staggered magnetization in the XXZ and TFI subsystems. (b)~Bipartition EE as a function of the bond index $n$, where $n=1$ denotes the end of the MPS and $n=15$ is the center; the inset shows the MPS structure used in this calculation. Data in this panel is for $L=10$ and $\chi_\mathrm{max}=1024$. (c)~QDL diagnostic for the effective entanglement between subsystems $A$ and $B$ shown in the inset. (d)~Staggered magnetization in the XXZ and TFI subsystems when the initial configuration is not an eigenstate of the time-averaged Hamiltonian. }
	\label{fig:DMRG_driven_composite}
\end{figure}

For a slow drive ($\omega=0.5$), the magnetization in both TFI and XXZ subsystems rapidly decays to half its initial value on the time scale of just a single period $T$.
Over the course of approximately 10 periods, the magnetization vanishes almost entirely, which is indicative of fast heating towards a trivial high-temperature paramagnetic state. 
In strong contrast, with the fast drive ($\omega=5$) we find signatures of a long-lived pre-thermal state which preserves the finite magnetization of the ground state configuration (Fig.~\ref{fig:DMRG_driven_composite}a). 
In both the TFI and XXZ subsystems a finite magnetization is maintained for more than 250 periods of the extensive subsystem drive, which is the maximum time duration in our numerical simulation. 

We substantiate the qualitative difference between the slow drive and the fast drive by considering two different entanglement diagnostics. 
First, we calculate the dynamics of the bipartite EE, which reveals that under the slow drive the system becomes maximally entangled such that the EE scales as $S_{\rm vN} = n \log 2$ in the long-time limit, where $n$ is the bond-index in the matrix product state (MPS) representation of the system (Fig.~\ref{fig:DMRG_driven_composite}b).
Such volume-law scaling of the EE is implied by a heating of the system to infinite temperature; note that the finite bond dimension $\chi_\mathrm{max}$ of the MPS causes a deviation from the scaling near the center $(n=15)$ of the chain when the EE approaches its theoretical upper bound $\log \chi_\mathrm{max} \approx 7$. 
In the fast-driven case, when $\omega = 5$, the bipartite EE plateaus far before reaching the upper bound and does not seem to follow the volume law.
It implies that the system does not thermalize within a moderate time scale, and the system instead remains in a pre-thermal phase. 
As a second entanglement measure we calculate the QDL diagnostic which provides a more detailed characterization of entanglement in a multi-component system. 
Unlike the bipartite EE, the QDL diagnostic allows us to extract the effective entanglement between two \emph{arbitrary} subsystems, which do not necessarily need to combine to the entire system. 
Let us consider $S_\mathrm{TFI-XXZ}$, which corresponds to the entanglement between the top XXZ chain and the TFI chain after a projective measurement on the bottom XXZ chain (see the definition of the QDL diagnostic in the Method section and the inset of Fig.~\ref{fig:DMRG_driven_composite}c.) 
For the slow drive, we observe a steep growth in the QDL diagnostic, indicating that spins on each chain -- which are close to a product state initially -- become strongly entangled in a short time (Fig.~\ref{fig:DMRG_driven_composite}c). 
In contrast, for the fast drive, the QDL diagnostic remains vanishingly small for the entire time scale observed, in line with our expectations for a pre-thermal regime. 

Finally, we briefly address the case when the initial configuration of the system is \emph{not} an eigenstate of the time-averaged Hamiltonian. 
To this end, we prepare the system in the ground state of the equilibrium model at $h(t)\equiv 0$ and subsequently apply the periodic drive. 
Generally, one would not expect a pre-thermal regime to arise. 
However, in analogy to the results discussed above, we still observe persistent magnetization for fast driving $\omega = 5$; the steady-state magnetization is only slightly reduced when compared to its initial value, see Fig.~\ref{fig:DMRG_driven_composite}d. 
We speculate that this rigidity against alteration of the initial configuration is a consequence of small variation of the ground state wave function on finite-sized systems within the ordered phase of our three-chain model; for an isolated TFI chain, which shows greater variation throughout the ordered phase phase, such rigidity is not observed (see SM).
We conclude that the environment of polarizable XXZ chains adds extra stability not only to the ground state magnetization of the embedded TFI chain, but also to its time evolution.


{\it Conclusion --}
In this letter, we ask whether an interacting quantum system can enter a long-lived pre-thermal steady state when quantum-quench protocols or time-dependent drives are applied only to an extensive subsystem. 
Using the DMRG and TDVP for time evolution, we study the example of two XXZ spin chains coupled to a TFI chain, where only the TFI chain is quantum-quenched or driven from a fully ordered ground state. 
In the case of sudden onset of the transverse field, when the strength of the transverse field is bigger than any other energy scale, it is shown that a pre-thermal steady state arises due to emergent quasi-conservation laws. 
In a similar spirit, when a sufficiently fast time-dependent drive of the transverse field is applied, the system develops a pre-thermal state where the magnetization remains finite across the system and the entanglement between the spin chains remains small. 
Our coupled spin-chain model is partly motivated by recent experiments on the organic material $\kappa$-H$_3$(Cat-EDT-TTF)$_2$ (`H-Cat') and its deuterated analogue. 
In those materials, layers of interacting electron systems (represented by XXZ spin chains in our model) are coupled via hydrogen bonds, where protons tunnel quantum mechanically in a double-well potential (spanned by bi-stable hydrogen bond configurations and modeled by transverse-field Ising spins in our setup) with an intrinsic time scale.
In first-principle calculations, the estimated tunnel barrier in H-Cat implies a tunneling rate of $10^{11}-10^{14}$~Hz~\cite{Yamamoto2016}. 
While the actual tunneling rate may be affected by the presence of other molecules attached to the hydrogen bond~\cite{Yamamoto2016}, it is conceivable that the phonon-assisted optic mode associated with the hydrogen tunneling would couple to infra-red light~\cite{Horsewill2008,Mochida1994}. 
It would thus be interesting to explore whether an external optical drive in the infrared regime can be utilized to study the dynamic properties of H-Cat and the possibility to stabilize a pre-thermal regime (in the sense of our extensive subsystem drive model) in these organic compounds. 
Further, if the tunneling rate of the hydrogen atoms can be tuned to be larger than all other scales, then even in the absence of any external driving, the system behaves as if it was being `self-driven' at a frequency given by the hydrogen tunneling rate. 
Therefore, if one prepares the system in the ground state of the symmetry-broken phase and evolves it with the time-independent Hamiltonian corresponding to a large tunneling rate, one still expects a pre-thermal symmetry-broken regime whose time now scales exponentially with the hydrogen tunneling rate.


\begin{acknowledgments}
The numerical simulations were performed on the Cedar cluster, hosted by WestGrid and Compute Canada and implemented with the TeNPy library~\cite{tenpy}. FLB and YBK were supported by the NSERC of Canada and the Center for Quantum Materials at the University of Toronto. HYL was supported by a National Research Foundation of Korea (Grant No. NRF-2020R1I1A3074769). TG acknowledges support by the National Science Foundation under Grant No. DMR-1752417, and by an Alfred P. Sloan Research Fellowship.
\end{acknowledgments}


\bibliography{trichain}

\end{document}


\title{Supplemental Material:\\Pre-thermalization via self-driving and external driving of extensive subsystems}

\author{Finn Lasse Buessen}
\thanks{These authors contributed equally to this work}
\affiliation{Department of Physics, University of Toronto, Toronto, Ontario M5S 1A7, Canada}

\author{Hyun-Yong Lee}
\thanks{These authors contributed equally to this work}
\affiliation{Department of Applied Physics, Graduate School, Korea University, Sejong 30019, Korea}
\affiliation{Division of Display and Semiconductor Physics, Korea University, Sejong 30019, Korea}
\affiliation{Interdisciplinary Program in E$\cdot$ICT-Culture-Sports Convergence, Korea University, Sejong 30019, Korea}

\author{Tarun Grover}
\affiliation{Department of Physics, University of California at San Diego, La Jolla, California 92093, USA}

\author{Yong Baek Kim}
\affiliation{Department of Physics, University of Toronto, Toronto, Ontario M5S 1A7, Canada}

\date{\today}
\maketitle


\begin{figure}
	\includegraphics[width=\linewidth]{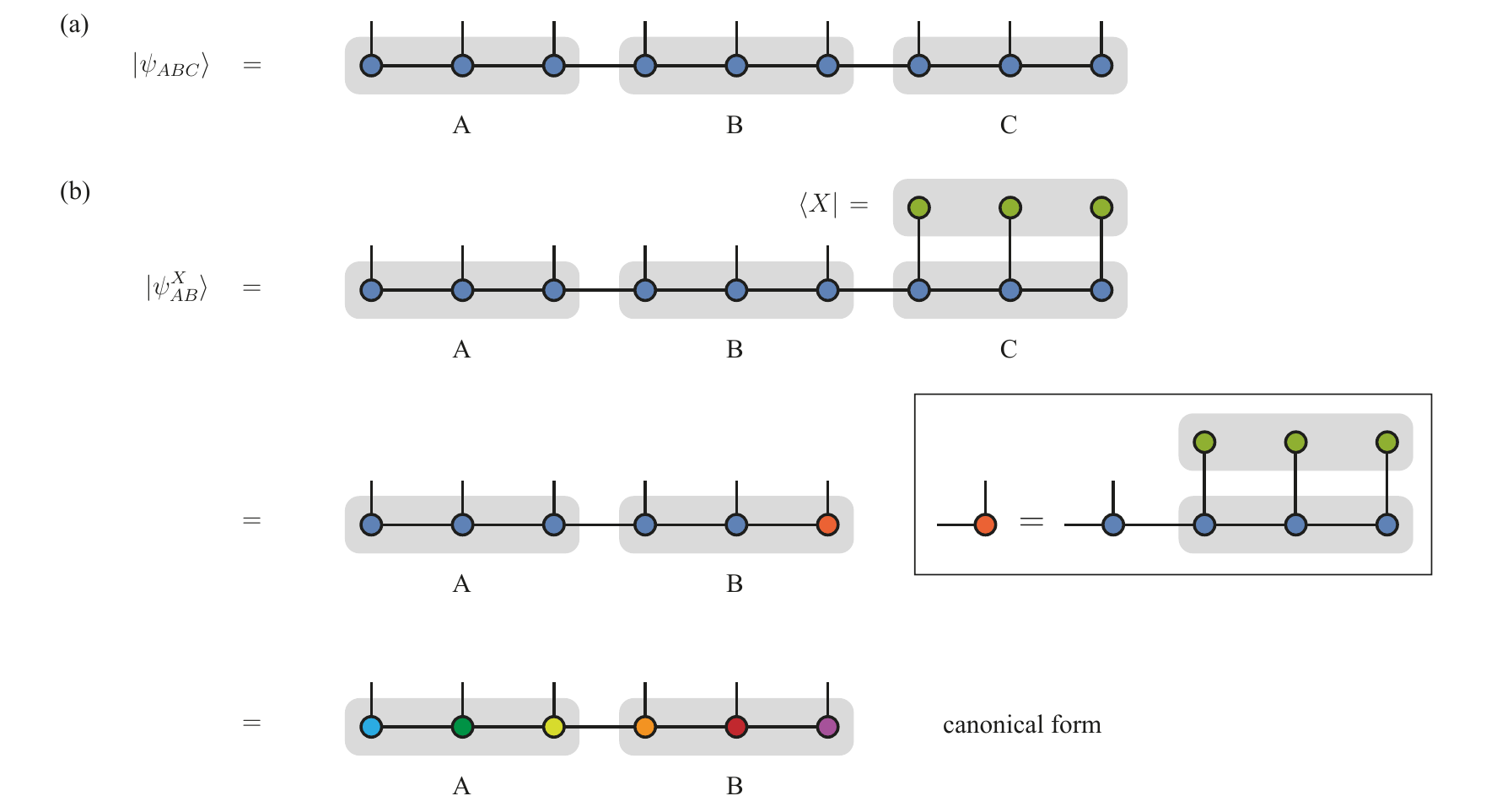}
	\caption{ {\bf MPS representations} of (a)~a possible tripartition of the system, and (b)~the projected state $|\psi_{AB}^X\rangle = \langle X| \psi_{ABC}\rangle $. }
	\label{fig:qdl_mps}
\end{figure}

\section{Numerical details on the QDL diagnostic}
\label{sec:QDLdiagnostic}
To measure a quantum disentangled liquid (QDL) diagnostic, we first divide a system $|\psi_{ABC}\rangle$ into three parts, say subsystems $A$, $B$, and $C$, as illustrated in Fig.~\ref{fig:qdl_mps}a. 
We then project the subsystem $C$ into a basis state $|X\rangle = \prod_{c\in C} |x_c\rangle$, with the $|x_c\rangle$ assuming values $|\uparrow\rangle$ or $|\downarrow\rangle$, resulting in a projected state $|\psi_{AB}^X\rangle = \langle X | \Psi_{ABC}\rangle$. 
Here, we assume that the whole wave function is normalized, i.e., $\langle \psi_{ABC}| \psi_{ABC}\rangle = 1$; hence, the projected state in general is no longer normalized. 
We then measure the norm of the projected state, $p_X = \langle \psi_{AB}^X | \psi_{AB}^X\rangle$, and the bipartite entanglement entropy $S_\mathrm{vN}^X$ between the subsystems $A$ and $B$.
Note that measuring $p_X$ and $S_\mathrm{vN}^X$ can be efficiently done by making use of the MPS representation: For a given configuration $|X\rangle$, the projection corresponds to contractions of physical bonds to local states, $|\uparrow\rangle$ or $|\downarrow\rangle$, see Fig.~\ref{fig:qdl_mps}b, and thus it does not require a many-body matrix-vector multiplication. 
After the projection, one can convert the projected state $|\psi_{AB}^X\rangle$ into the canonical form and evaluate the bipartite entanglement entropy between the subsystems $A$ and $B$ without calculating the density matrix $\rho_{AB}^X = |\psi_{AB}^X\rangle\langle \psi_{AB}^X|$ or the reduced density matrix $\rho_{A}^X = \mathrm{Tr}_{B}(|\psi_{AB}^X\rangle\langle \psi_{AB}^X|)$. 

However, the number of possible basis states $|X\rangle$ for the projection increases exponentially with the size of the subsystem~$C$. 
Thus, the exact computation of the QDL diagnostic, which is defined as $S_\mathrm{QDL} = \sum_{X} p_X S_\mathrm{vN}^X$ where the sum runs over all possible basis states $|X\rangle$, is exponentially hard. 
For instance, the exact computation of the QDL diagnostics $S_\mathrm{XXZ-half}$ and $S_\mathrm{TFI-half}$ considered in the main text requires $2^{2L}$ calculations of $p_X$ and $S_\mathrm{vN}^X$, where $L$ is the length of the spin chain. 
In order to mitigate the exponential complexity, we employ Monte Carlo sampling of the most significant basis states to approximate $S_\mathrm{vN}^X$. 
Specifically, we apply the Metropolis algorithm to create a Markov chain of basis states in the following way. 
First, we choose a random configuration for the basis state $|X\rangle$ and compute the associated $p_{X}$. 
Next, we flip the first spin in the subsystem $C$ to obtain a new configuration $|X'\rangle$, measure $p_{X'}$, and accept the configuration $|X'\rangle$ if $p_{X'}/p_{X}$ is larger than a random number, $\alpha$, which is uniformly distributed in $[0,1)$. 
If $p_{X'}/p_{X} < \alpha$, the configuration $|X'\rangle$ is rejected. 
We repeat the procedure for all spins in the subsystem C, and refer to the process of having attempted one update on every spin in subsystem C as one `sweep'. 
Every five sweeps we take a measurement of the bipartite entanglement entropy $S_\mathrm{vN}^X$, until $N=100$ measurements $s_1,\dots,s_{100}$ have been performed. 
We then estimate the QDL diagnostic from the statistical samples as $S_{\rm QDL} \approx \frac{1}{N} \sum_{ i=1 }^N s_i$.


\begin{figure}
	\includegraphics[width=\linewidth]{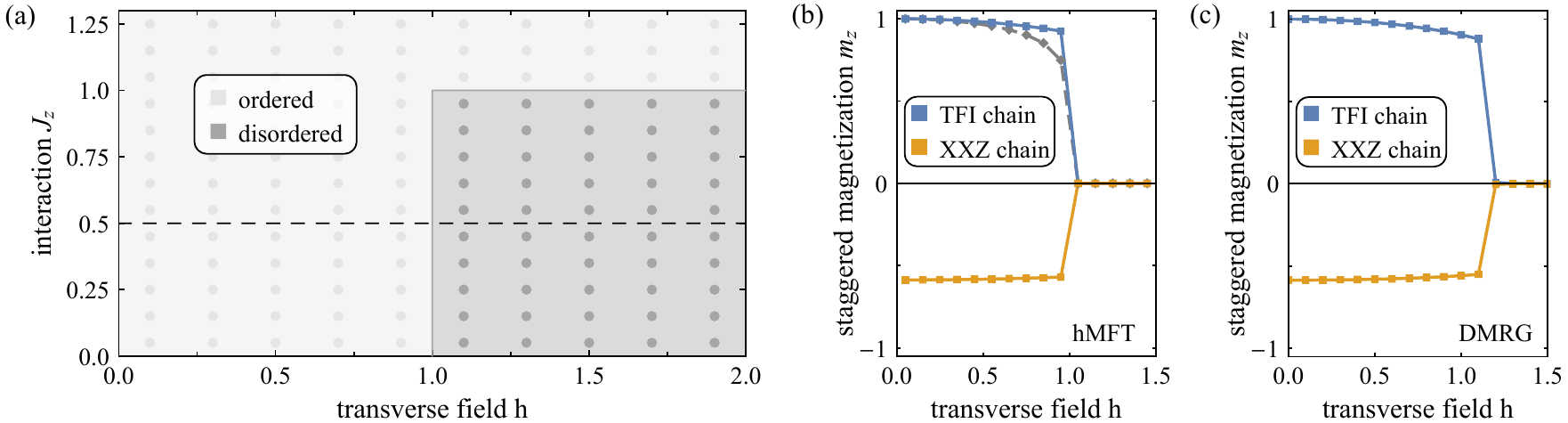}
	\caption{ {\bf Ground state phase diagram} of the three-chain model. (a)~Phase diagram as a function of transverse field $h$ and interaction $J_z$. (b)~Staggered magnetization as a function of $h$ at fixed $J_z=0.5$, computed within the hMFT approach. The dashed gray curve indicates the magnetization for an isolated TFI chain. (c)~Magnetization as a function of $h$ at fixed $J_z=0.5$, computed within DMRG.}
	\label{fig:phasediagram_composite}
\end{figure}

\section{Ground state phase diagram}
\label{sec:gsphasediagram}
In this section, we discuss the ground state phase diagram of the three-chain model as a function of $h\equiv h(t)$ and $J_z$. 
In order to reduce the computational cost, we compute the phase diagram within a hybrid mean-field theory (hMFT) approach. 
Starting from the initial model Hamiltonian Eq.~(1) in the main manuscript, 
\begin{equation}
H = H_\mathrm{TFI} + H_\mathrm{XXZ,t} + H_\mathrm{XXZ,b} + H_\mathrm{int} \,,
\end{equation}
we replace the local magnetization operators $\sigma_i^z$ and $\sigma_{i,\alpha}^z$ in the interaction term $H_\mathrm{int}$ by their mean fields $\langle \sigma_i^z \rangle$ and $\langle \sigma_{i,\alpha}^z \rangle$, respectively. 
This approach reduces the fully interacting two-dimensional problem to a system of three one-dimensional chains, 
\begin{equation}
\label{eq:gsphasediagram:eqsystem}
\left\{ \begin{array}{l}
H_{\mathrm{XXZ},t} + g \sum\limits_i \langle \sigma_i^z \rangle \sigma^z_{i,t} \\
H_\mathrm{TFI} + g \sum\limits_i \sigma_i^z \big( \langle \sigma^z_{i,t}\rangle - \langle \sigma^z_{i,b} \rangle \big) \\
H_{\mathrm{XXZ},b} - g \sum\limits_i \langle \sigma_i^z \rangle \sigma^z_{i,b} 
\end{array} \right. \,,
\end{equation}
that are coupled via their spatially inhomogeneous mean fields, which assume the role of an effective magnetic field $h_\mathrm{eff}$. 
This system of mean-field decoupled microscopic models then needs to be solved self-consistently.
In order to obtain the self-consistent solution, we solve each individual chain numerically at vanishing $h_\mathrm{eff}$ with the density matrix renormalization group (DMRG)~\cite{White1992,White1993} and follow an iterative approach: The initial solution from the DMRG calculation is used to calculate new estimates of $h_\mathrm{eff}$, which, in turn, are subsequently used as input to repeat the DMRG calculation. 
Eventually, a stable result for the mean-field configuration is obtained. 
By following the hMFT approach, the full quantum mechanical nature of each individual chain is retained, as their entanglement structure is faithfully resolved in the underlying DMRG calculation. 
In contrast, the inter-chain coupling does not mediate any entanglement; it loses its quantum mechanical character as a result of the mean-field decoupling. 
The hMFT approach is suitable to correctly resolve symmetry-broken states with finite magnetization (and typically low entanglement), but it cannot correctly resolve inter-chain entanglement or paramagnetic states (when the magnetization vanishes, the chains would effectively decouple.) 

In our implementation of the hMFT approach, we employ an infinite DMRG (iDMRG)~\cite{McCulloch2008} approach to obtain the ground state of the system within a periodic unit cell of $L=16$ sites, encoded in a matrix product state (MPS) representation with a maximum bond dimension of $\chi_\mathrm{max}=128$. 
The MPS is optimized until the energy density converges to an error smaller than $\Delta E_\mathrm{chain} = 10^{-8}$. 
In order to avoid vanishing expectation values of the magnetization even inside magnetically ordered phases due to the superposition of degenerate ground state configurations with opposite magnetization, we apply a weak (staggered) pinning field of magnitude $h_\mathrm{pin}=10^{-6}$ when optimizing the MPS. 
We then counteract an unphysical buildup of magnetic response as a consequence of the repeated application of pinning fields throughout the iterative solution of the hMFT equations by truncating the effective magnetic field for values smaller than $h_\mathrm{eff}=10^{-3}$. 
The iterative calculation is repeated until stationary values for $h_\mathrm{eff}$ are achieved and the energy difference between iterations falls below $\Delta E = 10^{-8}$, which is the precision to which we obtain the iDMRG solution for each individual chain. 

The resulting phase diagram is displayed in Fig.~\ref{fig:phasediagram_composite}a. 
As discussed in the main manuscript, we observe two distinct phases: (i)~a phase of composite magnetic order with finite magnetization in every chain, and (ii)~a disordered phase.  
The existence of only composite order or composite disorder can be conceptualized from the observation that both the TFI subsystem~\cite{Fogedby1978} and the XXZ subsystem~\cite{Alcarez1995,Kudasov2018} are unstable towards staggered longitudinal magnetic fields; a finite magnetization in one subsystem can thus be easily imprinted onto the other. 
Although we do not explore the whole phase diagram using the full three-chain DMRG, it is expected that one finds a qualitatively similar phase diagram. 
For instance, the magnetization as a function of $h$ obtained within the hMFT approach and the full DMRG are presented in Figs.~\ref{fig:phasediagram_composite}b and \ref{fig:phasediagram_composite}c, respectively. 
Both approaches identify the two different phases, i.e., fully ordered and fully disordered phases, and their discontinuous transition. 
Note that determining the precise phase boundary within the full DMRG approach requires a finite-size scaling analysis, which we do not perform here; the hMFT calculations are performed on an infinite chain, using iDMRG to obtain the self-consistent mean-field solution.


\begin{figure}
	\includegraphics[width=\linewidth]{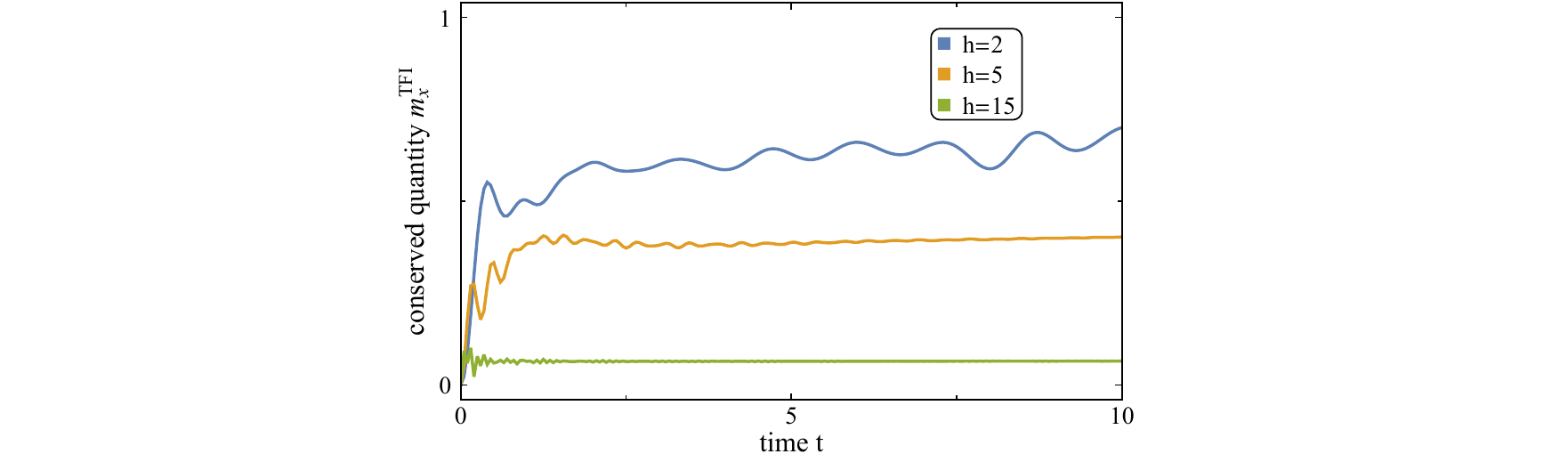}
	\caption{ {\bf Conserved quantity} $m_x^\mathrm{TFI}$ in the TFI chain after a quantum quench. Data shown is for chain length $L=30$ and $\chi_\mathrm{max}$=512. The conserved quantity is shown for three different values of the (post-quench) transverse field strength, $h=2$, $h=5$, and $h=15$. }
	\label{fig:DMRG_quench_conserved_composite}
\end{figure}

\section{Emergent conserved quantity after a quantum-quench}
\label{sec:conservation}
In the main manuscript, we have argued that after quantum-quenching the transverse field $h(t)$ in our coupled three-chain model to become the dominant energy scale, two conservation laws emerge. 
We discussed the conservation of $\chi^{yz}\equiv\frac{1}{L}\sum_{\langle i,j \rangle} (\sigma^y_i \sigma^y_j + \sigma^z_i \sigma^z_j)$ in the main manuscript; here, we show numerical evidence of the conservation of $m_x^\mathrm{TFI}\equiv \frac{1}{L} \sum_i \sigma_i^x$. 
While the transverse magnetization $m_x^\mathrm{TFI}$ exhibits significant non-trivial time dependence for weak post-quench transverse field $h=2$, increasing the field strength to $h=15$ leads to an almost constant time-evolution of the quantity, as expected (see Fig.~\ref{fig:DMRG_quench_conserved_composite}).


\begin{figure}
	\includegraphics[width=\linewidth]{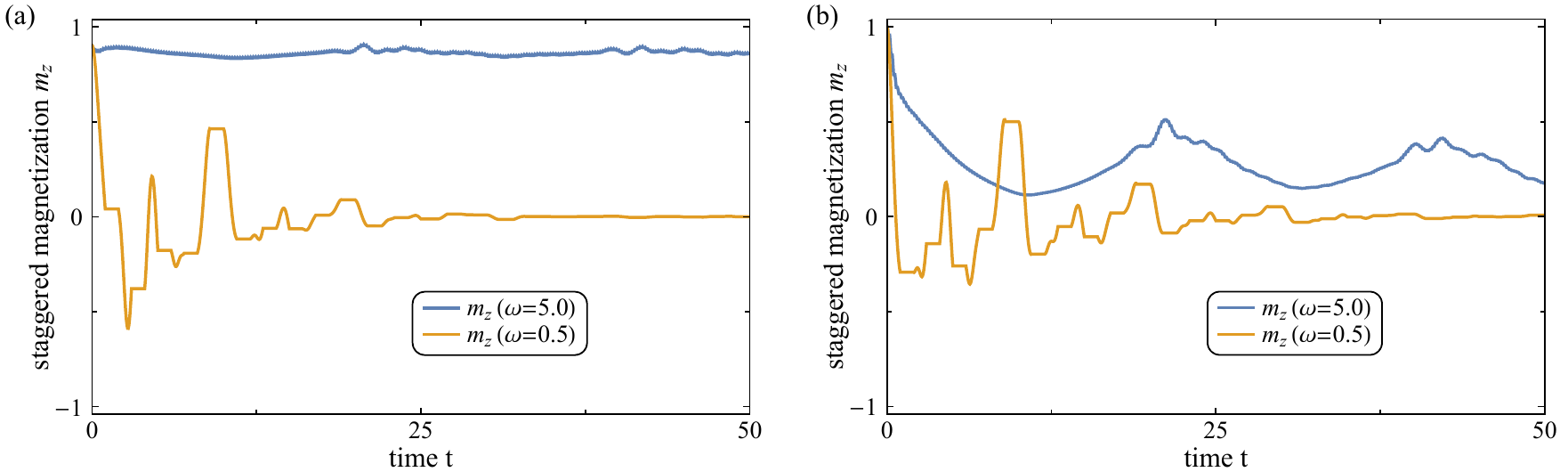}
	\caption{{\bf Transverse-field Ising chain} subject to periodic driving. The driving protocol is step-like, akin to the one we considered for the coupled three-chain model. The (staggered) magnetization $m_z$ is shown for either a fast drive $\omega=5$ or a slow drive $\omega=0.5$. Before applying the periodic drive, the TFI chain is initially prepared in the ground state of (a)~the time-averaged transverse field $h=0.75$ or (b)~at vanishing transverse field $h=0$. Data shown is for chain length $L=30$ and $\chi_\mathrm{max}=256$, in analogy to the parameters chosen for the three-chain model. }
	\label{fig:TFI:TFI_driven_magnetization_composite}
\end{figure}

\section{Periodically driven transverse field Ising chain}
\label{sec:TFI}
Our analysis of the coupled three-chain model, consisting of two XXZ spin chains which surround the central TFI spin chain, naturally raises the question of how the composite system behaves differently from just a single isolated TFI spin chain. 
For its equilibrium ground state, comparison with the phase diagram of the three-chain model seems to suggest a simple answer: The existence of the quantum phase transition at $h=1$ in an isolated TFI chain remains unaltered, as long as the XXZ spin chains are intrinsically disordered, i.e. $J_z<J_x$. 
Only if the XXZ spin chains are intrinsically prone to order, the quantum phase transition of the TFI chain vanishes and the composite system remains ordered even at $h>1$ as discussed in Sec.~\ref{sec:gsphasediagram}. 
In the parameter regime which is at the focus of our analysis (intrinsically disordered XXZ chains at $J_z/J_x = 0.5$) the impact of the embedding of the TFI chain into an environment of XXZ chains therefore appears negligible. 

However, the similarity of the equilibrium properties is misleading; the situation is not necessarily as simple when we compare the driven systems. 
In fact, we find that embedding the TFI spin chain into a polarizable (yet intrinsically disordered) environment of XXZ chains increases the stability of the composite magnetic order. 
This is seen as follows. 
We first prepare a TFI chain of $L=30$ sites in the ground state of the time-averaged field $h=0.75$. 
We further apply a weak symmetry-breaking staggered longitudinal magnetic field of strength $h_z=10^{-4}$ to obtain a symmetry-broken ground state which is comparable to the setting in our three-chain model. 
Next, we subject the system to a rapid external drive, which is defined analogously to the drive considered in the main manuscript. 
Confirming our general expectation for a rapidly driven system ($\omega=5$), we observe a prethermal regime during which a finite magnetization is maintained. 
If, on the other hand, the drive is slow ($\omega=0.5$), the magnetization rapidly diminishes (Fig.~\ref{fig:TFI:TFI_driven_magnetization_composite}a). 
We compare this finding to the scenario where the initial state is not the ground state of the time-averaged Hamiltonian. 
More precisely, we now initialize the model at vanishing transverse field $h=0$. 
With this setup, one expects that even at rapid periodic driving the system no longer necessarily enters a prethermal regime. 
Indeed, we observe that the magnetization now decays much faster even at rapid periodic driving, see Fig.~\ref{fig:TFI:TFI_driven_magnetization_composite}b. 

We interpret the numerical experiment discussed above by comparing it to analogous calculations for our three-chain model, which are shown in Figs.~3a,d in the main manuscript. 
From the direct comparison it is clear that the isolated TFI chain and the TFI chain embedded in an environment of XXZ chains behave similarly when their initial state is the ground state of the respective time-averaged Hamiltonian. 
In both cases, a prethermal regime manifests and the finite magnetization is maintained. 
In the alternate setting, however, when the initial state is no eigenstate of the time-averaged Hamiltonian, a discrepancy is observed: The diminishing of magnetization is much more pronounced in the isolated TFI chain. 
A possible explanation for the increased robustness of the magnetization in the three-chain model can be found in the ground-state phase diagram, where it can be seen that the magnetization shows only little variation within the ordered phase as a function of the transverse field strength $h$ (Fig.~\ref{fig:phasediagram_composite}b). 
We can therefore assume that the ground state of the zero-field model $h=0$ and that of the time-averaged Hamiltonian $h=0.75$ are similar, and performing the time evolution under either Hamiltonian leads to similar results. 
This is in contrast to the behavior of an isolated TFI chain where the variation of the ground state within the ordered phase is more pronounced~\cite{Pfeuty1970} (Fig.~\ref{fig:phasediagram_composite}b) and the evolution under the zero-field Hamiltonian $h=0$ deviates from the that of the time-averaged model $h=0.75$.


\bibliography{trichain}